\documentclass{article}

\usepackage{amsmath}
\usepackage{amssymb}
\usepackage[pdftex]{color}
\usepackage{graphicx}
\usepackage{epsfig}

\makeatletter
 \def\@biblabel#1{#1.}
\makeatother

\newcommand{\eref}[1]{(\ref{#1})}

\begin{document}

\begin{center}
{\Large\bf Crossovers in the non-Markovian dynamics of two-qubit
entanglements}\\*[3mm]
Shin-Tza Wu\\*[3mm]
{\em Department of Physics, National Chung Cheng University, Chiayi 621, Taiwan}\\
(Date: June 29, 2011)
\end{center}

\date{June 29, 2011}
\begin{quotation} \small
We study the entanglement dynamics of two non-interacting, spatially
separated qubits subject to local environment noises. Based on
exactly solvable models for non-Markovian amplitude damping and
phase damping noises, we are able to analyze the entanglement
dynamics of the two qubits for different coupling bandwidths and
different detunings. We show that entanglement oscillations can
occur for both amplitude and phase damping noises. Moreover, we
demonstrate that changing the coupling bandwidth can lead to
crossover between dissipative and non-dissipative entanglement
dynamics, while varying the detuning controls the crossover between
strong and weak coupling limits. Our findings can help provide a
synthesized picture for the entanglement dynamics of two qubits
subject to local environment noises.

\vspace*{5mm}


\end{quotation}


\section{Introduction}
Advancements of quantum information sciences during the past few
decades have been bringing new lights into researches on the
fundamental issues of quantum mechanics \cite{NC}. Among them,
studies of entanglement dynamics for open quantum systems have
occupied a central part due to the crucial role of quantum
entanglement in many aspects of quantum information sciences
\cite{BEZ}. For general multi-qubit systems, the study of
entanglement dynamics remains a major challenge due to the lack of a
manageable quantitative measure for entanglement in such systems
\cite{Horodecki}. In the case of two-qubit systems, however, there
have been reliable and easily computable entanglement measures which
have greatly assisted investigations of entanglement dynamics in
these systems. In particular, the degradation of entanglement in
two-qubit systems in the presence of environment noises has been an
issue of great theoretical and experimental interests \cite{YE09}.

In this paper, we shall examine the entanglement dynamics of two
spatially separated, non-interacting qubits subject to local
environment noises. This may correspond to, for instance, a pair of
entangled two-level atoms located in two different (lossy) cavities,
where each atom is separately coupled to the cavity modes. In order
to study the entanglement evolution of the qubits under the action
of environment noises, we will consider the cases of amplitude
damping noise, which can cause energy relaxations in the qubits, and
of phase damping noise, which can degrade the phase coherence of the
qubits \cite{NC}. The Markovian dynamics of the qubit entanglement
for such systems have been studied extensively \cite{YE09}. Novel
effects such as the complete destruction of entanglement within
finite time (the entanglement sudden death) have been predicted and
observed experimentally \cite{YE09,YE04,ESD1,ESD2}. There have also
been researches on the non-Markovian entanglement dynamics in such
systems. In particular, entanglement oscillations due to the finite
memory time of the environment coupling have been predicted
theoretically for amplitude damping noise \cite{Bellomo,Li10}, and
observed experimentally in photonic systems \cite{Xu}. In this
paper, based on analytically solvable models that exhibits
non-Markovian characteristics, we will show that for both amplitude
damping and phase damping noises, crossovers between oscillatory and
monotonic decaying entanglement evolution can occur in connection
with the transition between the non-dissipative and dissipative
limits of the environment coupling, which may also be regarded as a
crossover between the non-Markovian and the Markovian limits.
Furthermore, our model shall also enable us to examine the crossover
between the limits of weak and strong environment couplings.
Therefore, our results demonstrate not only that entanglement
oscillations can occur also for two qubits under phase damping
noises (cf.~Ref.~\cite{YE10}), but also offer detailed analysis for
the entanglement dynamics of two qubits in various regimes which can
help provide a synthesized view for the related problems. It should
be noted, however, that here our emphasis will lie in the
entanglement dynamics of the qubits, rather than the different
measure(s) for the ``non-Markovianity" of the dynamical maps that we
will discuss \cite{Wolf,Breuer,Rivas}; even though there are
connections between the two (see later). In other words, it is the
physical reasons behind the appearance/disappearance of entanglement
oscillations that will concern us, instead of any specific
quantitative measure for the ``non-Markovianity" of the dynamical
processes.

We will begin in the following section by a brief summary for our
theoretical framework. On the basis of Kraus representation
\cite{Kraus}, we shall explain how the dynamics of two independent
qubits can be constructed based on the single-qubit dynamics. Then
in Secs.~\ref{AD} and \ref{PD} we will apply this formulation to
examine the non-Markovian dynamics of two qubits subject to,
respectively, amplitude damping and phase damping noises. Finally,
Sec.~\ref{fin} provides a summary of our results and some further
discussions.

\section{Theoretical formulation}
\label{formulation}
Let us consider two spatially separated qubits that are interacting
with their local environments independently from each other. We shall
model the environment as a wide spectrum of harmonic oscillator modes so
that for each qubit the total Hamiltonian reads (we shall take
$\hbar=1$ throughout)
\begin{eqnarray}
H =  \frac{\omega_0}{2} \sigma_z + \sum_k \omega_k b_k^\dagger b_k + H_I \, .
\label{H_tot}
\end{eqnarray}
Here $\omega_0$ is the energy separation between the qubit levels
$|\pm\rangle$ (which have energies $\pm \frac{\omega_0}{2}$,
respectively), $\sigma_z$ is the third Pauli matrix, and $b_k$ is
the annihilation operator for the oscillator mode with frequency
$\omega_k$ in the environment degrees of freedom. The last term
$H_I$ in \eref{H_tot} denotes the interaction Hamiltonian for the
qubit and its local environment, whose explicit form will be
specified later. Since the two qubits are not interacting with each
other and their environment noises are uncorrelated, the two-qubit
dynamics can be constructed from the single-qubit dynamics based on
the Kraus formulation, as we shall now explain \cite{YE04}.

For the single-qubit Hamiltonian \eref{H_tot}, suppose one can
find the time evolution for the reduced density matrix
$\tilde{\rho}$, it is then possible to express \cite{Kraus}
\begin{eqnarray}
\tilde{\rho}(t) = \sum_{i=1}^2 E_i \tilde{\rho}(0) E_i^\dagger \, ,
\label{K_rho_1}
\end{eqnarray}
where $\tilde{\rho}(0)$ is the initial density matrix for the qubit
and $E_i$ are the operation elements associated with the time
evolution of the qubit. These operation elements are encoded with
effects of the environment noises on the qubit and satisfy
$\sum_{i=1}^2 E_i^\dagger E_i=1$ \cite{Kraus}. Once the single-qubit
operation elements are known, the time evolution of two independent
qubits $A$, $B$ can then be obtained from the operation elements
\cite{YE04}
\begin{eqnarray}
K_1 = E_1^A \otimes E_1^B \, ,
\nonumber\\
K_2 = E_1^A \otimes E_2^B \, ,
\nonumber\\
K_3 = E_2^A \otimes E_1^B \, ,
\nonumber\\
K_4 = E_2^A \otimes E_2^B \, ,
\label{Ks}
\end{eqnarray}
where the superscripts $A$, $B$ are the qubit labels. The reduced
density matrix $\rho$ for the two qubits thus evolves according to
\begin{eqnarray}
\rho(t) = \sum_{i=1}^4 K_i \rho(0) K_i^\dagger \,
\label{K_rho_2}
\end{eqnarray}
with $\rho(0)$ the initial density matrix for the two qubits.
Applying this formulation, in the following sections we shall
consider specific models for the system-environment interaction
$H_I$ for which the operation elements can be found explicitly. We
will consider two types of environment noises separately: the
amplitude damping noise, which can cause energy relaxations in the
qubits, and the phase damping noise, which can lead to dephasing in
the qubits \cite{NC}. Based on these results, we shall study the
time evolution of the two-qubit entanglement.

\section{Amplitude damping}
\label{AD}
\subsection{Single-qubit dynamics}
As our first case for the qubit-environment coupling, let us
consider a single qubit interacting with the environment by
absorbing and emitting energy quanta from the oscillator modes, and
at the same time making transitions between the upper and lower
qubit levels. This interaction would thus change the level
occupations in the qubit and cause energy relaxations (thus
frequently termed the amplitude damping channel). We model this
coupling by using the following interaction Hamiltonian in
\eref{H_tot} \cite{BP}
\begin{eqnarray}
H_I = \sum_k \left( g_k \sigma_- b_k^\dagger + g_k^* \sigma_+ b_k \right)
\, ,
\label{HI_AD}
\end{eqnarray}
where $g_k$ are the coupling constants and $\sigma_\pm$ are the
raising/lowering operators for the qubit levels. The time evolution
of the qubit in the presence of the coupling \eref{HI_AD} can be
solved exactly \cite{Berry} and the corresponding operation elements
in the basis $\{|+\rangle,|-\rangle\}$ are
\begin{eqnarray}
E_1 = \left(
           \begin{array}{cc}
                 p  &  0 \\
                 0  &  1
           \end{array}
      \right) \, ,
      \quad
E_2 = \left(
           \begin{array}{cc}
                 0  &  0 \\
                 q  &  0
           \end{array}
      \right) \, ,
\label{E_AD}
\end{eqnarray}
where $q\equiv\sqrt{1-|p|^2}$. Here $p$ is the solution for the
equation
\begin{eqnarray}
\frac{d}{d t}p(t) = - \int_0^t d\tau f(t-\tau) p(\tau)
\label{p_eq}
\end{eqnarray}
with $f$ the noise correlation function \cite{BP}
\begin{eqnarray}
&& f(t-\tau) \nonumber \\
&\equiv&
\left\langle \left( \sum_{k} g_k^* b_k e^{-i\omega_k t} \right)
\left(\sum_{k'} g_{k'} b_{k'}^\dagger e^{i\omega_{k'} \tau} \right) \right\rangle
\, e^{i\omega_0(t-\tau)}
\nonumber \\
&=& \sum_{k,k'} g_k^* g_{k'}
\langle b_k b_{k'}^\dagger \rangle \, e^{i(\omega_0-\omega_k)t}\,
e^{-i(\omega_0-\omega_{k'})\tau} \, . \label{f}
\end{eqnarray}
The expectation values in \eref{f} are taken with respect to the
state of the environment. For our calculation, we shall consider a
vacuum initial environment state. Thus the noise correlation
function \eref{f} reduces to
\begin{eqnarray}
f(t-\tau) &=& \sum_{k} |g_k|^2 \, e^{i(\omega_0-\omega_k)(t-\tau)}
\nonumber\\
&=& \int_{-\infty}^\infty d\omega\, J(\omega)\, e^{i(\omega_0-\omega)(t-\tau)}
\label{fJ}
\end{eqnarray}
with $J(\omega)$ the spectral function for the coupling \cite{BP}.
In the following, we will consider a specific form for the spectral
function that shall enable analytic solution for \eref{p_eq}. The
single qubit dynamics can thus be attained explicitly.

Let us consider a Lorentzian spectral function for the
qubit-environment coupling \cite{Breuer}
\begin{eqnarray}
J(\omega) = \frac{1}{2\pi} \frac{\gamma\lambda^2}{(\omega-\omega_0+\Delta)^2+\lambda^2}
\, .
\label{J_Lorz}
\end{eqnarray}
Here $\gamma$ is the (phenomenological) decay rate for the upper
qubit level $|+\rangle$, $\lambda$ is the bandwidth of the coupling
(thus $\lambda^{-1}$ characterizes the environment coherence time),
and $\Delta$ is the detuning from the resonance frequency
$\omega_0$. Utilizing \eref{J_Lorz} in \eref{fJ}, one can obtain the
correlation function $f$ easily and then solve for $p(t)$ from
\eref{p_eq}. One finds \cite{Li10}
\begin{eqnarray}
p(t) = e^{-\left(\frac{\lambda-i\Delta}{2}\right)t}
\left[\cosh\left(\frac{dt}{2}\right)+\frac{\lambda-i\Delta}{d}\sinh\left(\frac{dt}{2}\right)\right],
\label{p_AD}
\end{eqnarray}
where
\begin{eqnarray}
d\equiv\sqrt{(\lambda-i\Delta)^2-2\gamma\lambda}
\, .
\label{d}
\end{eqnarray}
With the explicit form for the function $p$, one can construct the
operation elements in \eref{E_AD}. Following the prescriptions of
Sec.~\ref{formulation} we are now ready for studying the two-qubit
dynamics.

\subsection{Two-qubit dynamics}
Let us now consider two non-interacting qubits that are coupled to
independent amplitude damping noises separately, each described by
\eref{HI_AD}. With the results for single qubit dynamics in the
preceding subsection, as described in Sec.\ref{formulation} one can
obtain the two-qubit operation elements $K_i$ using \eref{E_AD} and
\eref{p_AD} in \eref{Ks}. The time evolution of the reduced density
matrix for the two qubits then follows from \eref{K_rho_2}. Using
the basis $\{|++\rangle,|+-\rangle,|-+\rangle,|--\rangle\}$, we find
the matrix elements \cite{Li10}
\begin{eqnarray}
\rho_{11}(t) &=& |p_A\,p_B|^2 \rho_{11}(0) \, ,
\nonumber\\
\rho_{22}(t) &=& |p_A|^2 \left(\rho_{22}(0) + q_B^2\,\rho_{11}(0) \right)\, ,
\nonumber\\
\rho_{33}(t) &=& |p_B|^2 \left(\rho_{33}(0) + q_A^2\,\rho_{11}(0) \right)\, ,
\nonumber\\
\rho_{44}(t) &=& 1 - \left(\rho_{11}(t) + \rho_{22}(t) + \rho_{33}(t) \right)\, ,
\nonumber\\
\rho_{12}(t) &=& |p_A|^2 \,p_B \, \rho_{12}(0) \, ,
\nonumber\\
\rho_{13}(t) &=& p_A\,|p_B|^2 \, \rho_{13}(0) \, ,
\nonumber\\
\rho_{14}(t) &=& p_A\,p_B \, \rho_{14}(0) \, ,
\nonumber\\
\rho_{23}(t) &=& p_A\,p_B^* \, \rho_{23}(0) \, ,
\nonumber\\
\rho_{24}(t) &=& p_A \left(\rho_{24}(0) + q_B^2\,\rho_{13}(0) \right)\, ,
\nonumber\\
\rho_{34}(t) &=& p_B \left(\rho_{34}(0) + q_A^2\,\rho_{12}(0) \right)\, .
\label{rho_t_AD}
\end{eqnarray}
Here $p_\alpha$ (with $\alpha=A,B$ the qubit labels) are given by
\eref{p_AD} but with the parameters now carrying the qubit labels,
ie.~$\gamma\rightarrow\gamma_\alpha$,
$\lambda\rightarrow\lambda_\alpha$, $\Delta\rightarrow\Delta_\alpha$
(hence also $d\rightarrow d_\alpha$), and as in \eref{E_AD}, we
denote $q_\alpha\equiv\sqrt{1-|p_\alpha|^2}$.

In order to study the entanglement dynamics of the pair of qubits,
we shall adopt concurrence as the entanglement measure
\cite{Wootters}, namely
\begin{eqnarray}
C(t) = \max \Big\{\, 0 , \sqrt{\mu_1}-\sqrt{\mu_2}-\sqrt{\mu_3}-\sqrt{\mu_4} \Big\}
\, ,
\label{Ct}
\end{eqnarray}
where $\mu_i$ ($i=1\sim 4$) are eigenvalues of the matrix
$\rho(\sigma_y^A\otimes\sigma_y^B)\rho^*(\sigma_y^A\otimes\sigma_y^B)$
in descending orders; $\sigma_y^\alpha$ is the second Pauli matrix
for qubit $\alpha$. Using \eref{rho_t_AD} and \eref{p_AD}, one can
find the concurrence evolution for any initial density matrix
accordingly. For instance, for an initial density matrix of the
``X-form" \cite{YE_X}
\begin{eqnarray}
\rho(0) = \left(
                \begin{array}{cccc}
                      \rho_{11} &     0    &      0     & \rho_{14} \\
                          0     & \rho_{22} & \rho_{23} &     0     \\
                          0     & \rho_{32} & \rho_{33} &     0     \\
                      \rho_{41} &  0 & 0 & \rho_{44}
                \end{array}
          \right) \, ,
\label{rho_X}
\end{eqnarray}
it follows from \eref{rho_t_AD} and \eref{Ct} that the concurrence
for the reduced density matrix $\rho(t)$ is given by
\begin{eqnarray}
C(t) = 2 \max \Big\{\, 0 , & |\rho_{14}(t)|-\sqrt{\rho_{22}(t)\rho_{33}(t)} \,,
\nonumber\\
& |\rho_{23}(t)|-\sqrt{\rho_{11}(t)\rho_{44}(t)} \, \Big\}
\, .
\label{C_X}
\end{eqnarray}
In the following, we shall present results based on the calculations
outlined above. We will consider symmetric configurations in which
the two qubits are identical, so that they have the same level
separation $\omega_0$ and the same decay rate
$\gamma_A=\gamma_B=\gamma$, and are subject to amplitude noises of
identical characteristics, thus $\lambda_A=\lambda_B=\lambda$,
$\Delta_A=\Delta_B=\Delta$, and consequently $p_A=p_B=p$ in
\eref{rho_t_AD}.

\begin{figure}
\begin{center}
\includegraphics*[width=80mm]{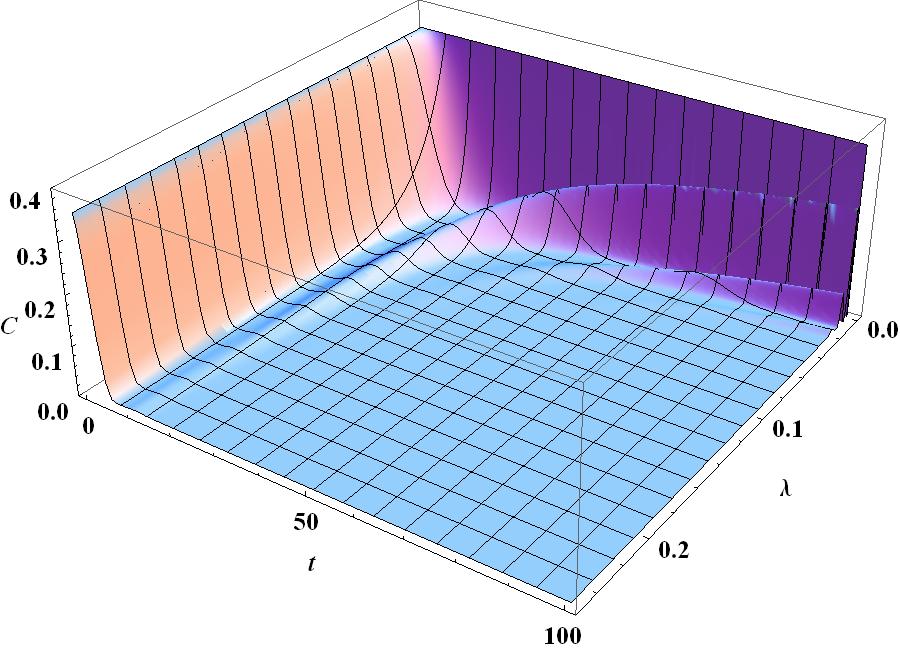}
\\*[-2mm]  (a) \\*[2mm]
\includegraphics*[width=80mm]{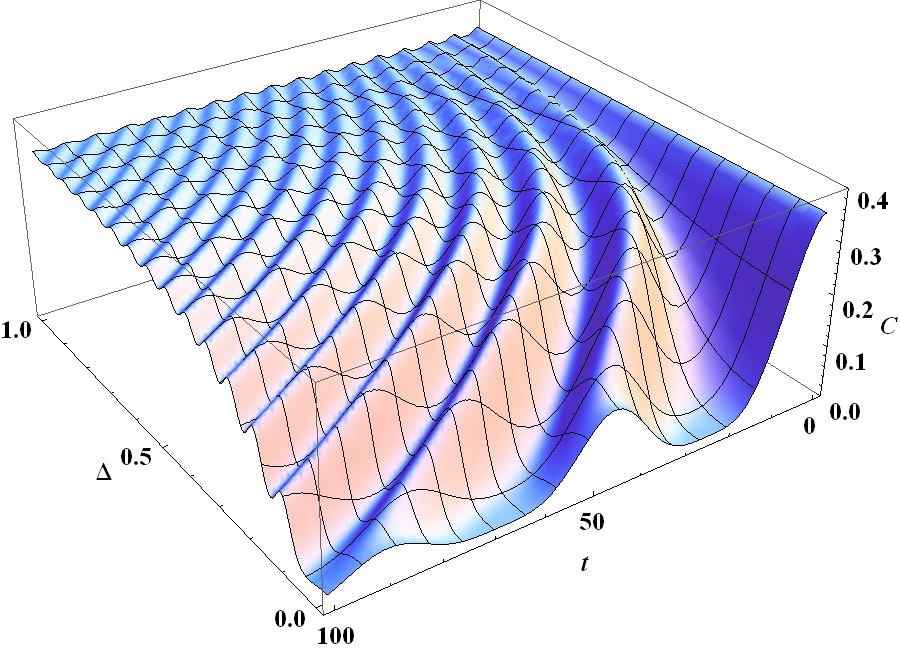}
\\*[-2mm]  (b) \\
\includegraphics*[width=80mm]{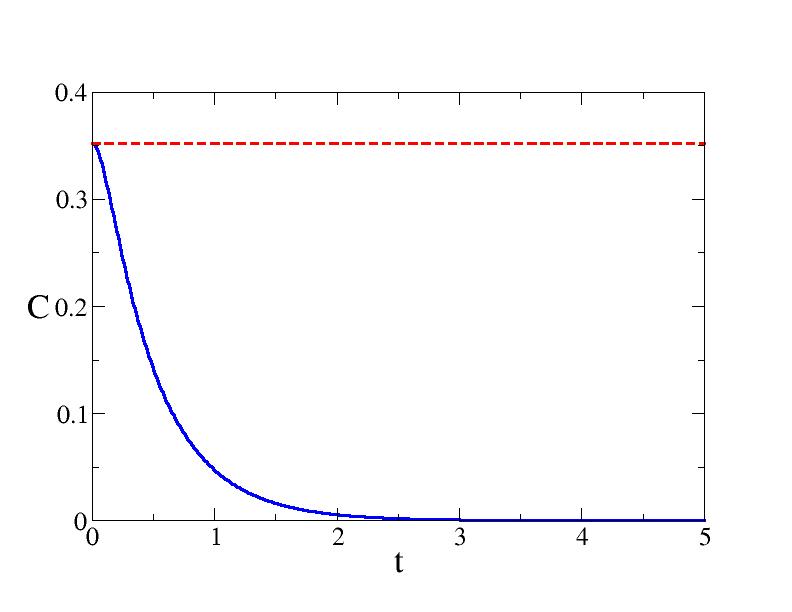}
\\  (c)
\end{center}
\caption{\small Entanglement evolution of two qubits with the
initial density matrix \eref{rho_X1} subject to amplitude damping noises.
We plot the concurrence for (a) different $\lambda$ at fixed detuning
$\Delta=0.01\gamma$ and (b) different $\Delta$ at fixed coupling bandwidth
$\lambda=0.01\gamma$. In (c) the solid curve depicts the entanglement
evolution for large coupling bandwidth $\lambda=10.0\gamma$
(with $\Delta=0.01\gamma$), and the dashed curve for large detuning $\Delta=10.0\gamma$
(with $\lambda=0.01\gamma$); note that for clarity, the time axis in (c) has a smaller range
than those in (a), (b). The axes $\lambda$ in (a) and $\Delta$ in (b) are in units of
$\gamma$, while the time axes in all plots are in units of $1/\gamma$.
\label{AD_fig1} }
\end{figure}

Fig.~\ref{AD_fig1} displays our results for the entanglement
evolution of two qubits with the initial density matrix
\begin{eqnarray}
\rho(0) = \frac{1}{3}
          \left(
                \begin{array}{cccc}
                      \frac{1}{3} &  0 & 0 &     0     \\
                            0     &  1 & 1 &     0     \\
                            0     &  1 & 1 &     0     \\
                            0     &  0 & 0 & \frac{2}{3}
                \end{array}
          \right) \, ,
\label{rho_X1}
\end{eqnarray}
which is an entangled mixed state. In Fig.~\ref{AD_fig1}(a) we plot
the concurrence evolution at fixed detuning $\Delta=0.01\gamma$ for
different values of $\lambda$. It is seen clearly that the crossover
from oscillatory behavior to monotonic decay occurs when the
coupling bandwidth changes from small to large values. In
Fig.~\ref{AD_fig1}(b) the concurrence evolution is plotted at fixed
coupling bandwidth $\lambda=0.01\gamma$ for different values of
detuning $\Delta$. One can observe that the entanglement evolution
changes from damped oscillatory behavior to barely damped one as the
detuning varies from small to large values. Since entanglement
preservation is a key issue in quantum information sciences, it is
of great interest to analyze in detail how such transitions arise.
Let us first attempt with simple, qualitative reasoning.

For the results in Fig.~\ref{AD_fig1}(a), when the coupling
bandwidth is small ($\lambda \ll \gamma$), it corresponds to the
limit with each qubit coupled to almost just one single oscillator
mode. The entanglement dynamics is thus non-dissipative since the
entanglement can transfer back and forth between the qubits and the
oscillator modes involved \cite{Yonac06,Yonac07}. In other words, in
the limit of long noise coherence time (ie.~large $\lambda^{-1}$)
the environment memory can reconstruct the qubit entanglement, so
that the concurrence displays an oscillation that is hardly damped.
In the other limit of large coupling bandwidth ($\lambda \gtrsim
\gamma$; see the solid line in Fig.~\ref{AD_fig1}(c)), the qubits
are coupled to a wide spectrum of oscillator modes. Thus the
environment memory time becomes exceedingly short, so that the
entanglement dynamics becomes Markovian and the concurrence decays
to zero monotonically. In the intermediate regime, due to the finite
environment memory time, concurrence oscillation is damped but
approaches zero in a non-monotonic manner. Since entanglement
oscillation is an indication for non-Markovian dynamics
\cite{Bellomo,Rivas,Li10}, the crossover associated with the
coupling bandwidth can also be regarded as one between the limits of
Markovian and ``extremely" non-Markovian dynamics.

For the results in Fig.~\ref{AD_fig1}(b), we note that for large
detuning ($\Delta\gtrsim\gamma$; see also the dashed line in
Fig.~\ref{AD_fig1}(c)) the qubits can only couple weakly to the
environment noises. The concurrence is thus barely damped in this
limit. With decreasing detuning, the environment noise can couple
more and more effectively to the qubits, which results in stronger
and stronger damping in the concurrence evolution. Therefore, the
crossover from undamped to damped oscillatory entanglement dynamics
that we are seeing in Fig.~\ref{AD_fig1}(b) is connected with the
transition between the weak and strong coupling limits in the
qubit-environment interaction.

In order to substantiate the arguments above, let us now turn to
more quantitative, detailed analysis for the behavior of $p(t)$
based on \eref{p_AD}.

\subsubsection{Small coupling bandwidth}
\label{SCB}
In the limit of small $\lambda$, suppose $\lambda \ll \Delta \ll
\gamma$, one can find from \eref{d} that
\begin{eqnarray}
d \simeq i\sqrt{2\gamma\lambda + \Delta^2} \equiv i\,\omega_d
\, .
\label{d_limit1}
\end{eqnarray}
Hence it follows from \eref{p_AD}
\begin{eqnarray}
p(t) \simeq e^{i\frac{\Delta}{2} t}
\left[ \cos\left(\frac{\omega_d}{2}t\right)
-\frac{i\Delta}{\omega_d}
\sin\left(\frac{\omega_d}{2}t\right) \right] \, ,
\label{p_AD_limit1}
\end{eqnarray}
which oscillates in time with no damping. This gives rise to the
undamped concurrence oscillations, which are manifestations of
non-dissipative dynamics. Note that this regime is small in
Fig.~\ref{AD_fig1}(a); it can be seen more clearly in
Fig.~\ref{AD_fig1}(b). For larger bandwidths $\Delta \ll \lambda \ll
\gamma$, one can obtain from \eref{d}
\begin{eqnarray}
d \simeq i\sqrt{2\gamma\lambda} \, ,
\label{d_limit2}
\end{eqnarray}
and thus \eref{p_AD} would yield
\begin{eqnarray}
p(t) \simeq e^{-\frac{\lambda t}{2}}
\cos\left(\sqrt{\frac{\gamma\lambda}{2}}t\right)
\, .
\label{p_AD_limit2}
\end{eqnarray}
We see that $p(t)$ is now oscillating with frequency
$\sqrt{\gamma\lambda/2}$ with a damping time $\sim \lambda^{-1}$.
Therefore for $\lambda \ll \gamma$, there can be many oscillations
within the damping time; one would thus find weakly damped
concurrence oscillations in this regime.

It should be noted that in the limit under consideration, if the
detuning is large, such that $\Delta\gg\gamma\gg\lambda$, then we
have from \eref{d}
\begin{eqnarray}
d \simeq \lambda-i\Delta \, .
\label{d_limit3}
\end{eqnarray}
It then follows readily from \eref{p_AD}
\begin{eqnarray}
p(t) \simeq 1
\, .
\label{p_AD_limit3}
\end{eqnarray}
The concurrence thus remains a constant throughout the time
evolution; see the dashed curve in Fig.~\ref{AD_fig1}(c). As
explained above, this is because the detuning is so large here that
the qubit-environment coupling becomes no longer effective. The
qubit dynamics thus becomes entirely immune to the environment
noise.

\subsubsection{Large coupling bandwidth}
\label{LCB}
In the limit of large coupling bandwidth, suppose
$\lambda\gg\gamma\gg\Delta$, it follows from \eref{d} that
\begin{eqnarray}
d \simeq \lambda - \gamma - i \Delta \, .
\label{d_limit4}
\end{eqnarray}
One can thus obtain from \eref{p_AD}
\begin{eqnarray}
p(t) \simeq e^{- \frac{\gamma}{2} t} \, ,
\label{p_AD_limit4}
\end{eqnarray}
which is typical Markovian behavior. The concurrence evolution is
therefore strongly damped in this regime and decays to zero
monotonically (see the solid curve in Fig.~\ref{AD_fig1}(c)). Note
that in arriving at \eref{p_AD_limit4} the detuning has been taken
to be small, ie.~$\Delta\ll \lambda, \gamma$. If we have large
detuning, so that $\lambda\gg\Delta\gg\gamma$ or
$\Delta\gg\lambda\gg\gamma$, it is easy to show that one would get
$p(t)\simeq 1$ in both cases. The concurrence thus becomes
stationary due to the ineffective coupling at large detunings.

\subsubsection{Spectral function}
To further justify the foregoing analysis in {\it 1.} and {\it 2.},
it is also instructive to examine the spectral function $J(\omega)$
in the corresponding limits. In the limit of small coupling
bandwidth, one has from \eref{J_Lorz}
\begin{eqnarray}
\lim_{\lambda\rightarrow 0} J(\omega) = \frac{\gamma\lambda}{2} \delta(\omega-\omega_0+\Delta)
\, ,
\label{J_limit1}
\end{eqnarray}
which signifies that, as noted above, each qubit is coupled to a
single oscillator mode with frequency $\omega_0-\Delta$. As a
consequence, a non-dissipative, oscillatory concurrence evolution
follows. In fact, making use of \eref{J_limit1} in \eref{fJ}, and
then solving $p(t)$ from \eref{p_eq}, one can recover
\eref{p_AD_limit1} accordingly. In the limit of large coupling
bandwidth, one finds similarly from \eref{J_Lorz}
\begin{eqnarray}
\lim_{\lambda\rightarrow \infty} J(\omega) = \frac{\gamma}{2\pi}
\, .
\label{J_limit2}
\end{eqnarray}
Therefore, we have a white noise spectrum for the qubit-environment
coupling in this limit. The entanglement dynamics thus becomes
Markovian and the concurrence decreases to zero monotonically.
Again, using \eref{J_limit2} in \eref{fJ}, one can regain
\eref{p_AD_limit4} easily from \eref{p_eq}. In the intermediate
regime between the two extremes, the concurrence displays damped
oscillations, where the oscillations are manifestations of the
finite environment memory times due to non-zero coupling bandwidths.

\begin{figure}
\begin{center}
\includegraphics*[width=80mm]{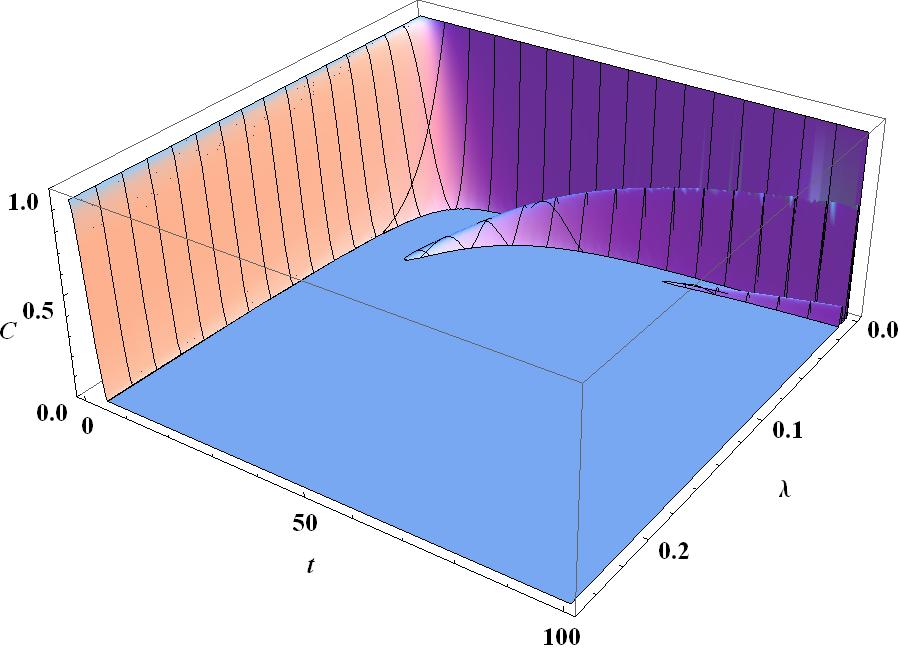}
\\*[-2mm]  (a) \\*[2mm]
\includegraphics*[width=80mm]{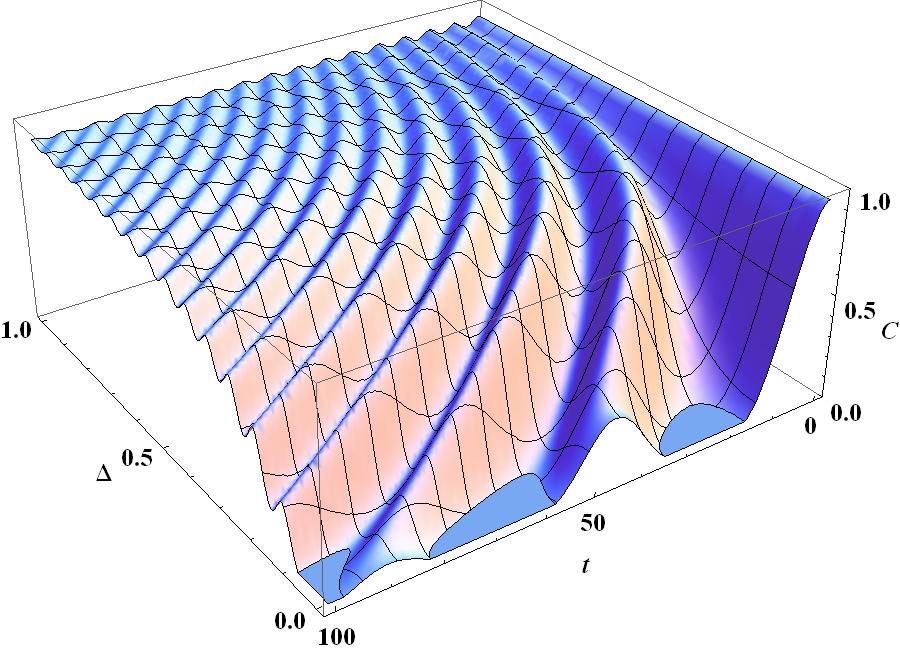}
\\*[-2mm]  (b) \\
\includegraphics*[width=80mm]{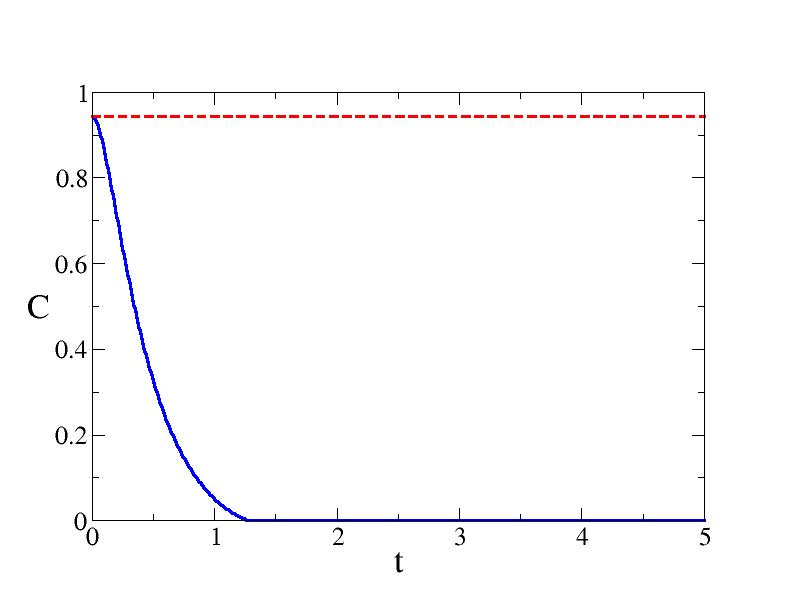}
\\  (c) \\
\end{center}
\caption{\small Same as Fig.~\ref{AD_fig1} but for the initial
density matrix \eref{rho_X2}. The flat regions at the base planes
of (a) and (b) are signatures of entanglement sudden death.
\label{AD_fig2} }
\end{figure}

As another example, we consider an entangled pure initial state
which has the density matrix
\begin{eqnarray}
\rho(0) =\frac{1}{3}
         \left(
                \begin{array}{cccc}
                        2       & 0 & 0 &  \sqrt{2} \\
                        0       & 0 & 0 &    0      \\
                        0       & 0 & 0 &    0      \\
                      \sqrt{2}  & 0 & 0 &    1
                \end{array}
          \right) \, .
\label{rho_X2}
\end{eqnarray}
The results for the calculation are shown in Fig.~\ref{AD_fig2}.
Although there are now regions with entanglement sudden death (and
rebirth), our explanations above for the concurrence
oscillation/damping in Fig.~\ref{AD_fig1} can be carried over.

\section{Phase damping}
\label{PD}
\subsection{Single-qubit dynamics}
Let us now turn to another class of qubit-environment interaction
that couples longitudinally to the (pseudo)spin of the qubit. The
environment noise would therefore not cause any energy relaxation in
the qubit, but randomize the relative phase between the qubit
levels. In the single-qubit Hamiltonian \eref{H_tot}, this coupling
can be modeled with the following qubit-environment interaction
\cite{Palma}
\begin{eqnarray}
H_I = \sum_k \sigma_z \left( g_k b_k^\dagger + g_k^*  b_k \right)
\, ,
\label{HI_PD}
\end{eqnarray}
where, as previously, $g_k$ are the coupling constants. Unlike
\eref{HI_AD}, here $H_I$ commutes with the bare qubit Hamiltonian
$(\omega_0/2)\sigma_z$, and thus would not induce any population
transfer between the qubit levels $|+\rangle$ and $|-\rangle$. The
coupling \eref{HI_PD} would instead act like a fluctuating magnetic
field along $z$-direction that can randomize the phase of the qubit
\cite{Palma}. This qubit-environment coupling is thus often referred
to as the phase damping channel \cite{NC}.

For the interaction Hamiltonian \eref{HI_PD}, one can work out the
time evolution of the qubit explicitly \cite{Palma}. It is then not
difficult to find the corresponding single-qubit operation elements.
Suppose the environment is initially in the vacuum state, in the
usual basis $\{|+\rangle,|-\rangle\}$ one can find
\begin{eqnarray}
E_1 = \left(
           \begin{array}{cc}
                 p &  0 \\
                 0 &  1
           \end{array}
      \right) \, ,
      \quad
E_2 = \left(
           \begin{array}{cc}
                 q  &  0 \\
                 0  &  0
           \end{array}
      \right) \, ,
\label{E_PD}
\end{eqnarray}
where, as before, $q\equiv\sqrt{1-|p|^2}$. Here we have
\begin{eqnarray}
p(t) = \exp\{\Gamma(t)\}
\label{p_PD}
\end{eqnarray}
with
\begin{eqnarray}
\Gamma(t) \equiv -2 \sum_k |g_k|^2 \int_0^t d\tau \int_0^t d\tau' e^{-i\omega_k(\tau-\tau')}
\label{Gamma}
\end{eqnarray}
a real negative function of time. As in the case of amplitude
damping, we can rewrite $\Gamma(t)$ of \eref{Gamma} in terms of the
noise correlation function $f(t)$ and the spectral function
$J(\omega)$ using \eref{fJ}; we get
\begin{eqnarray}
\Gamma(t) &=& -2 \int_0^t d\tau \int_0^t d\tau' f(\tau-\tau') e^{-i\omega_0(\tau-\tau')}
\nonumber \\
&=& -2 \int_0^t d\tau \int_0^t d\tau' \int_{-\infty}^\infty \!\! d\omega \, J(\omega) e^{-i\omega(\tau-\tau')}
\, .
\label{Gamma_J}
\end{eqnarray}
As previously, once a specific form for $J(\omega)$ is given, it is
then possible to find the qubit dynamics accordingly.

Let us consider again the Lorentzian spectral function
\eref{J_Lorz}. Note, however, that since the interaction
\eref{HI_PD} would not cause any transition between the qubit levels
$|+\rangle$ and $|-\rangle$, the frequency $\omega_0$ no longer
plays the role of the ``resonant frequency". One can thus redefine
$\omega_c \equiv \omega_0 - \Delta$ as the ``central frequency" for
the coupling (this follows from the form of the spectral function
\eref{J_Lorz}, which centers at $\omega=\omega_c$). Applying
\eref{J_Lorz} in \eref{Gamma_J}, one can obtain
\begin{eqnarray}
\Gamma(t) &=& -2 \gamma\lambda \Bigg\{
\frac{\lambda}{\lambda^2+\omega_c^2} t
\nonumber \\
&&- \frac{1}{(\lambda^2+\omega_c^2)^2} \Big[ (\lambda^2 -
\omega_c^2) \left( 1 - e^{-\lambda t} \cos(\omega_c t) \right)
\nonumber \\
&&+ 2 \lambda \omega_c e^{-\lambda t} \sin(\omega_c t) \Big] \Bigg\}
\, . \label{Gamma_t}
\end{eqnarray}
It is interesting to note that for non-zero $\omega_c$, $\Gamma(t)$
(and hence $p(t)$) can oscillate in time. The qubit decoherence can
consequently occur non-monotonically. In the case of two qubits,
this can lead to entanglement oscillations, as we shall now show.

\subsection{Two-qubit dynamics}
For two independent qubits $A$, $B$ subject to uncorrelated local
phase noises, again, we resort to the prescription outlined in
Sec.~\ref{formulation}. Using \eref{E_PD} in \eref{Ks}, one can
derive the two-qubit operation elements in the phase damping
channel. Substituting the results into \eref{K_rho_2}, one can then
obtain the time evolution of the the matrix elements for the
two-qubit reduced density matrix (in the basis
$\{|++\rangle,|+-\rangle,|-+\rangle,|--\rangle\}$)
\begin{eqnarray}
\rho_{ii}(t) &=& \rho_{ii}(0) \, , \quad \mbox{$i=1 \sim 4$.}
\nonumber\\
\rho_{12}(t) &=& p_B \, \rho_{12}(0) \, ,
\nonumber\\
\rho_{13}(t) &=& p_A \, \rho_{13}(0) \, ,
\nonumber\\
\rho_{14}(t) &=& p_A\,p_B \, \rho_{14}(0) \, ,
\nonumber\\
\rho_{23}(t) &=& p_A\,p_B \, \rho_{23}(0) \, ,
\nonumber\\
\rho_{24}(t) &=& p_A \, \rho_{24}(0) \, ,
\nonumber\\
\rho_{34}(t) &=& p_B \, \rho_{34}(0) \, ,
\label{rho_t_PD}
\end{eqnarray}
where $p_\alpha$ are the expressions for qubit $\alpha=A,B$ in
accordance with \eref{p_PD} and \eref{Gamma_t} (with appropriate
qubit labels added). Note that here we have made use of the fact
that $p_{A,B}(t)$ are real functions of time. From \eref{rho_t_PD}
we see that, as noted previously the phase noise does not affect the
qubit occupations since the diagonal elements of the reduced density
matrix remain unchanged. However, it diminishes the coherence of the
density matrix by corroding the off-diagonal elements of the reduced
density matrix.

In order to examine the entanglement dynamics ensued from
\eref{rho_t_PD}, again we shall consider initial density matrices of
the X-form \eref{rho_X}. The concurrence of the qubits can then be
calculated using \eref{p_PD}, \eref{Gamma_t} together with
\eref{rho_t_PD} in \eref{C_X}. As previously, we shall consider
symmetric configurations where the two qubits are identical (so that
they have the same level separation $\omega_0$ and the same decay
rate $\gamma_A=\gamma_B=\gamma$) and the phase noises acting on them
have the same characteristics, namely $\lambda_A=\lambda_B=\lambda$,
$\omega_c^A=\omega_c^B=\omega_c$, and consequently $p_A=p_B=p$ in
\eref{rho_t_PD}.

\begin{figure}
\begin{center}
\includegraphics*[width=80mm]{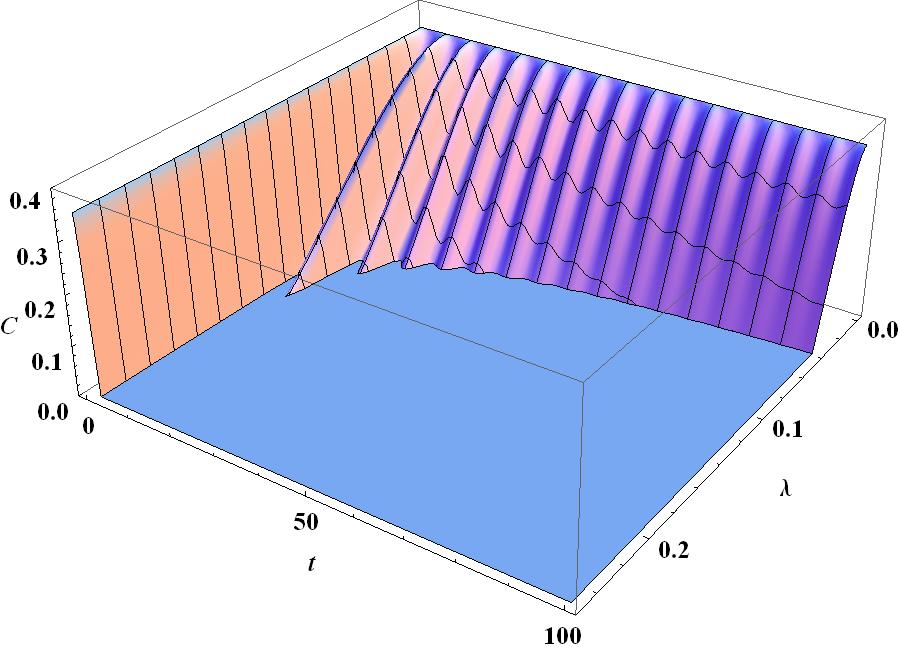}
\\*[-2mm]  (a) \\*[2mm]
\includegraphics*[width=80mm]{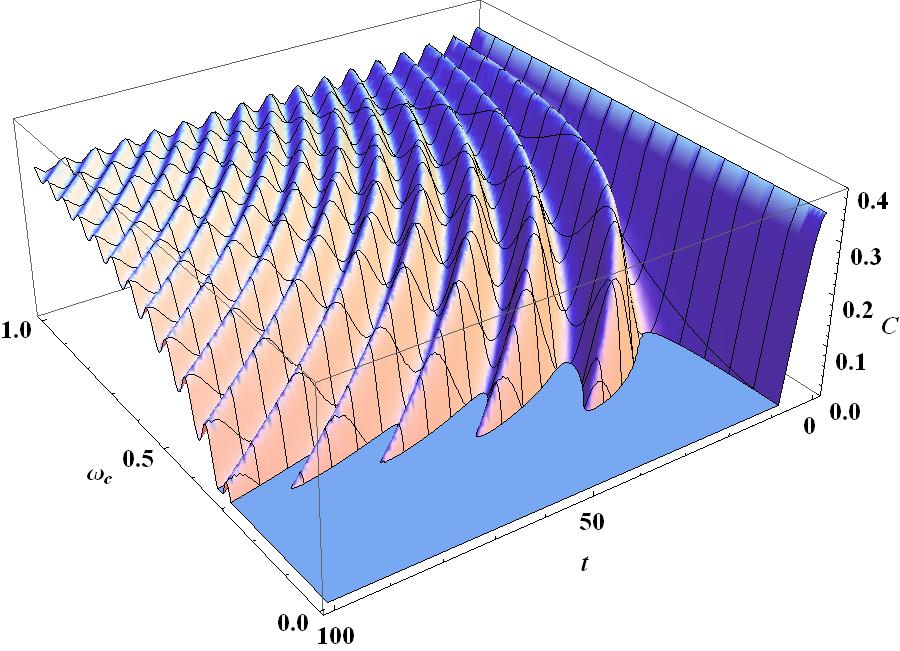}
\\*[-2mm]  (b) \\
\includegraphics*[width=80mm]{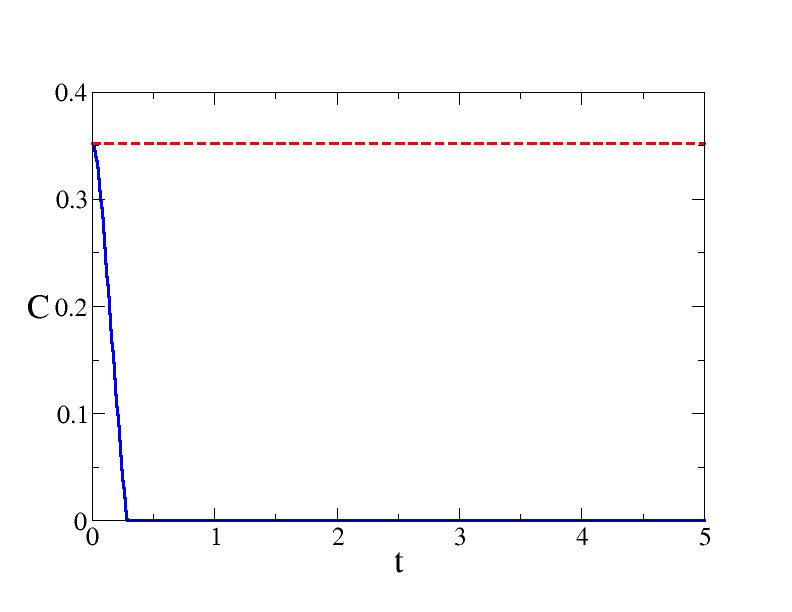}
\\  (c) \\
\end{center}
\caption{\small Entanglement evolution of two qubits with the
initial density matrix \eref{rho_X1} subject to phase damping noises. Here
we plot the concurrence for (a) different $\lambda$ at fixed central frequency
$\omega_c=1.0\gamma$ and (b) different $\omega_c$ at fixed coupling bandwidth
$\lambda=0.01\gamma$. In (c) the solid curve illustrates
the entanglement evolution for large coupling bandwidth $\lambda=10.0\gamma$
(with $\omega_c=1.0\gamma$), and the dashed curve for large central frequency
$\omega_c=10.0\gamma$ (with $\lambda=0.01\gamma$). For clarity, we plot in
(c) a smaller range of time than in (a), (b). The units for the
axes are the same as in Fig.~\ref{AD_fig1}. The flat areas at the base
planes in (a) and (b) indicate regions with entanglement sudden death.
\label{PD_fig1}
}
\end{figure}

\begin{figure}
\begin{center}
\includegraphics*[width=80mm]{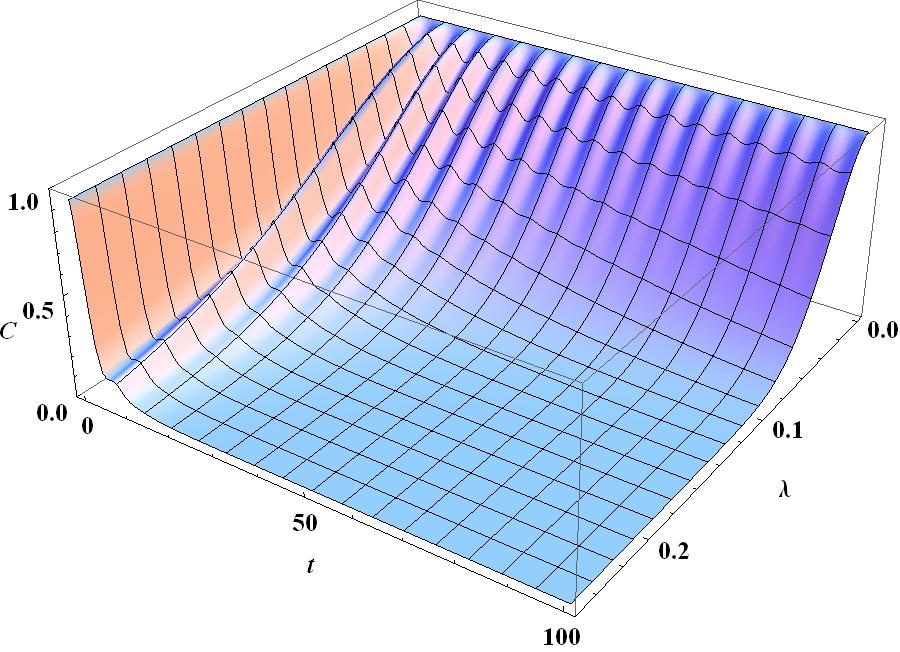}
\\*[-2mm]  (a) \\*[2mm]
\includegraphics*[width=80mm]{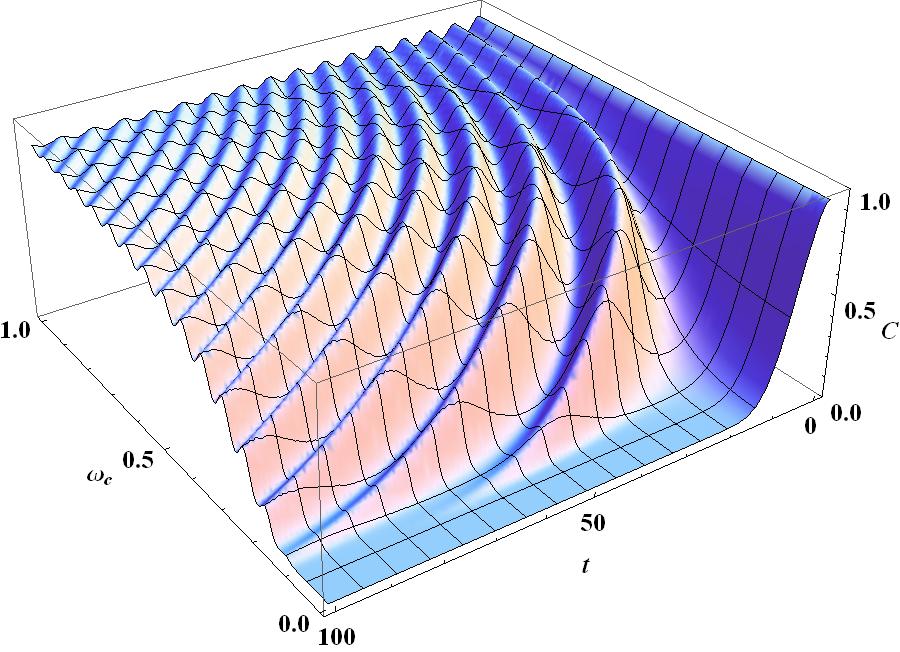}
\\*[-2mm] (b) \\
\includegraphics*[width=80mm]{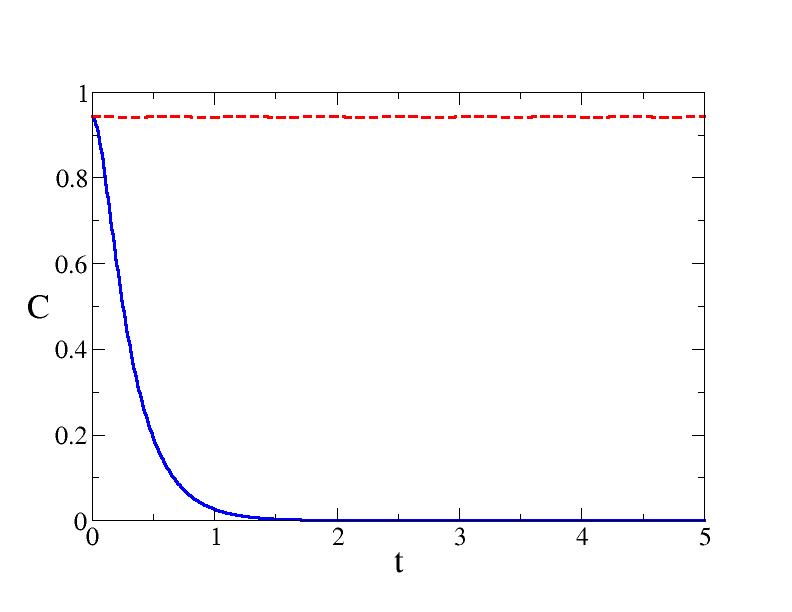}
\\  (c) \\
\end{center}
\caption{\small Same as Fig.~\ref{PD_fig1} but for the initial
density matrix \eref{rho_X2}.
\label{PD_fig2} }
\end{figure}

Fig.~\ref{PD_fig1} illustrates our results for the entanglement
evolution of two qubits with the initial density matrix
\eref{rho_X1}. In Fig.~\ref{PD_fig1}(a) we plot the concurrence
evolution at fixed central frequency $\omega_c=1.0\gamma$ for
different coupling bandwidths $\lambda$. When the coupling bandwidth
varies from small to large values, crossover from oscillatory
behavior to monotonic decay in the concurrence evolution can be seen
clearly. In Fig.~\ref{PD_fig1}(b) the concurrence evolution is
plotted at fixed coupling bandwidth $\lambda=0.01\gamma$ for
different central frequencies $\omega_c$. One can observe that the
entanglement evolution changes from monotonically decaying behavior
to barely damped oscillations as $\omega_c$ varies from small to
large values. Another set of results for the entanglement dynamics
under phase damping noise is illustrated in Fig.~\ref{PD_fig2},
where we consider the initial density matrix \eref{rho_X2}. Again,
similar crossovers between different regimes are clearly visible.

Qualitatively, the results in Figs.~\ref{PD_fig1} and \ref{PD_fig2}
can be understood in the same manner as presented in Sec.~\ref{AD}
for the entanglement evolution under amplitude damping noise. For
instance, in both Figs.~\ref{PD_fig1}(a) and \ref{PD_fig2}(a) the
regime with small coupling bandwidths correspond to the limit with
each qubit coupled to one single oscillator mode. The entanglement
dynamics thus exhibits non-dissipative behavior, as entanglement
swapping can take place between the pair of qubits and the two
oscillator modes. The undamped entanglement oscillations in this
regime are consequences of the long environment memory time
$\lambda^{-1}$, which enables entanglement reconstructions in the
qubits. For large coupling bandwidths, the wide spectrum of the
qubit-environment coupling entails an extremely short environment
memory time. Therefore, the Markovian entanglement dynamics emerges
and the concurrence decays to zero monotonically (see the solid
curves in Figs.~\ref{PD_fig1}(c) and \ref{PD_fig2}(c)). For the
intermediate regime, the finite coupling bandwidth implies
non-vanishing environment memory time. The concurrence thus carries
out damped oscillations, in which the oscillatory part reflects the
memory effect that survives the phase damping noise.

As pointed out earlier in the present section, the central frequency
$\omega_c$ is the peak frequency for the qubit-environment coupling.
Therefore, at large $\omega_c$ the qubits are coupled to the phase
noise only very weakly. This is why in both Figs.~\ref{PD_fig1}(b)
and \ref{PD_fig2}(b) the concurrence is barely damped at large
$\omega_c$ (see also the dashed curves in Figs.~\ref{PD_fig1}(c) and
\ref{PD_fig2}(c)). With decreasing $\omega_c$, as the noise can
couple more and more effectively to the qubits, damping in the
concurrence evolution gradually sets in. Eventually, in the limit of
small $\omega_c$, the entanglement oscillation disappears and the
concurrence drops to zero monotonically.

As in Sec.~\ref{AD}, more quantitative analysis for the results here
can be achieved by looking into the behavior of $p(t)$ according to
\eref{p_PD} and \eref{Gamma_t} in the limits of small and large
coupling bandwidths. It should be noted, however, that unlike
amplitude damping noise, here the regimes of small and large
coupling bandwidths are determined from the relative magnitude
between $\lambda$ and $\omega_c$ (instead of the qubit decay rate
$\gamma$). This is because under the action of phase damping noise,
the qubit populations would not change. Thus the time scale $\gamma$
becomes inessential to the non-Markovian dynamics of the qubit (see
later). Instead, it is the central frequency $\omega_c$ that comes
into play (which is clear from the explicit form of $\Gamma(t)$ in
\eref{Gamma_t}). Therefore, for phase damping noise although the
central frequency $\omega_c$ plays a role similar to the detuning
$\Delta$ in amplitude damping noise, the concurrence oscillation
depends more sensitively on $\omega_c$ (compare
Figs.~\ref{AD_fig1}(b) with \ref{PD_fig1}(b), and
Figs.~\ref{AD_fig2}(b) with \ref{PD_fig2}(b), especially for regions
with small $\Delta$ and $\omega_c$). With the above observations,
let us now examine $p(t)$ for small and large coupling bandwidths
using \eref{p_PD} and \eref{Gamma_t}.

\subsubsection{Small coupling bandwidth}
For narrow coupling bandwidths $\lambda \ll \omega_c$, we find from
\eref{Gamma_t} that
\begin{eqnarray}
\Gamma(t) \simeq - \frac{2\gamma\lambda}{\omega_c^2}
\left[1 + \lambda t - e^{-\lambda t} \cos(\omega_c t) \right]
\, .
\label{Gamma_t_limit1}
\end{eqnarray}
Therefore for times $t \ll \lambda^{-1}$, $p(t)=\exp\{\Gamma(t)\}$
oscillates with negligible damping and the concurrence dynamics
exhibits non-dissipative characteristics. This can again be
anticipated from the spectral function $J(\omega)$, which reduces to
the form for single mode coupling in this limit (see
\eref{J_limit1}). Indeed, using the spectral function
\eref{J_limit1} in \eref{Gamma_J}, one can derive an expression for
$\Gamma(t)$ which has exactly the same form as \eref{Gamma_t_limit1}
except with $\lambda t=0$ (ie.~being totally non-dissipative). For
$\lambda t \sim 1$, however, we see from \eref{Gamma_t_limit1} that
damping begins to set in. As a result, the oscillating cosine term
in \eref{Gamma_t_limit1}, which represents the environment memory
effect, is attenuated and a damped concurrence oscillation thus
follows. At long times $t\gg \lambda^{-1}$, one can infer from
\eref{Gamma_t_limit1} that $p(t)$ eventually drops to zero
exponentially, thus also the concurrence (save the complications due
to entanglement sudden death, such as in Fig.~\ref{PD_fig1}).

Note that in the present limit, if the central frequency is so large
that $\omega_c \gg \gamma, \lambda$, it follows from
\eref{Gamma_t_limit1} that $\Gamma(t)$ would become vanishing. In
this case we have $p(t)\simeq 1$ and the entanglement becomes
stationary, as the qubit is completely decoupled from the
environment in this limit. This can be seen from the dashed curves
in Figs.~\ref{PD_fig1}(c) and \ref{PD_fig2}(c).

\subsubsection{Large coupling bandwidth}
For large coupling bandwidths $\lambda \gg \omega_c$, it follows
from \eref{Gamma_t} that
\begin{eqnarray}
\Gamma(t) \simeq -2 \gamma \left( t - \frac{1 - e^{-\lambda t} \cos(\omega_c t)}{\lambda} \right)\, .
\label{Gamma_t_limit2}
\end{eqnarray}
Since $\lambda$ is now large, the damping time (or environment
memory time) $\lambda^{-1}$ is short. Within the damping time, the
oscillatory cosine term in \eref{Gamma_t_limit2} cannot even
complete one period since $\omega_c \lambda^{-1} \ll 1$ here; the
environment memory is therefore washed away quickly. For times $t
\gg \lambda^{-1}$, we get from \eref{Gamma_t_limit2} that
\begin{eqnarray}
p(t) = \exp\{\Gamma(t)\} \simeq e^{-2\gamma t} \, ,
\label{p_PD_limit2}
\end{eqnarray}
which manifests the Markovian nature of this limit. As a
consequence, the concurrence becomes strongly damped and exhibits
Markovian behavior \cite{YE10}. As previously, we can further
corroborate the argument above by considering the spectral function
$J(\omega)$ in the present limit; the white noise spectrum (see
Eq.~\eref{J_limit2}) is again responsible for the Markovian
entanglement dynamics in this limit. As a check, one can use
\eref{J_limit2} in \eref{Gamma_J}, which would yield
\eref{p_PD_limit2} immediately.

\section{Discussions and Conclusions}
\label{fin}
In summary, based on exactly solvable models, we have studied the
entanglement dynamics of a pair of non-interacting qubits that are
separately coupled to their local environments. In particular, we
examine in detail the crossover between the limits of dissipative
and non-dissipative entanglement dynamics that are manifested in the
damped/undamped concurrence oscillations. We show that this
crossover is connected with the bandwidth of the qubit-environment
coupling. We have also studied the crossover between the limits of
strong and weak couplings, which is controlled by the detuning of
the qubit-environment interaction. Both categories of crossovers
have been considered for two types of environment noises, the
amplitude damping channel and the phase damping channel, and
physical pictures for the results have been provided.

Although we have illustrated our results only for two specific
initial density matrices \eref{rho_X1} and \eref{rho_X2}, our
conclusions are indeed fairly general. For instance, suppose only
one of the qubits is coupled to the environment, it is easy to
derive from our formulas that the concurrence evolution for any pure
initial state would follow the factorization law \cite{fact,Li10}
\begin{eqnarray}
C(t) = |p(t)| C(0) \, ,
\label{fact_law}
\end{eqnarray}
where $p(t)$ is given by \eref{p_AD} and \eref{p_PD}, respectively,
for amplitude and phase damping noises. Since for maximally
entangled pure initial states $C(0)=1$, it follows that $p(t)$
completely determines the concurrence evolution of the the maximally
entangled state for this noisy channel. Therefore, according to
\eref{fact_law}, for any pure initial state the concurrence evolves
according to the action of the noisy channel over the maximally
entangled state \cite{fact}. The oscillation/damping of $p(t)$ in
different regimes would thus exhibit also in the concurrence
evolution for general pure initial states. Similar results exist
also for certain class of mixed initial states when both qubits are
subject to local noises \cite{Li10}.

An issue that we did not address in this paper is the interplay
between amplitude noise and phase noise. For Markovian entanglement
dynamics, Yu and Eberly had demonstrated that the combination of
phase damping and amplitude damping noises can give rise to
non-additive entanglement dissipation \cite{YE06}. For the
non-Markovian case, it is certainly of interest to investigate
whether additional effects would emerge when both phase and
amplitude noises are present. Also, in our analysis of the
entanglement evolution, we did not examine the detailed dynamics
close to entanglement sudden death. We wish to take up these issues
in future works.

\section*{Acknowledgments}
I would like to thank Prof. Sungkit Yip and Mr Tseng-Jen Huang for
valuable discussions. This research is supported by NSC of Taiwan
through grant nos. NSC 96-2112-M-194-011-MY3 and NSC 99-2112-M-194
-009 -MY3; it is also partly supported by the Center for Theoretical
Sciences, Taiwan.


\end{document}